\documentclass[a4paper,11pt]{article}
\pdfoutput=1 

\usepackage{jheppub} 

\usepackage[T1]{fontenc} 

\usepackage{graphicx}
\usepackage{subfig}
\usepackage[dvipsnames]{xcolor}

\usepackage{booktabs}

\usepackage{xltabular}

\newcolumntype{C}{>{$}c<{$}}

\newcolumntype{s}{>{\hsize=.6\hsize}X}

\newcolumntype{K}{>{\centering\arraybackslash}X}

\newcolumntype{Y}{>{\arraybackslash}X}

\usepackage{diagbox}

\usepackage{multirow}
\usepackage{hyperref}

\usepackage{colortbl}

\def\[{\left[}
\def\]{\right]}
\def\({\left(}
\def\){\right)}

\newcommand{\beq}{\begin{equation}}
\newcommand{\eeq}{\end{equation}}
\newcommand\beqa{\begin{eqnarray}}
\newcommand\eeqa{\end{eqnarray}}

\begin{document}

\title{\boldmath Conformal field theory-data analysis for $\mathcal{N} = 4$ Super-Yang-Mills at strong coupling}

\author[s]{Julius Julius}
\author[t]{and Nika Sokolova}

\affiliation[s]{Harish-Chandra Research Institute, Homi Bhabha National Institute, Chhatnag Road, Jhunsi, Allahabad 211019, India}
\affiliation[t]{Department of Mathematics, King's College London, Strand WC2R 2LS, United Kingdom}

\emailAdd{julius@hri.res.in}
\emailAdd{nika.sokolova@kcl.ac.uk}

\abstract{
We analyse the CFT-data of planar 4D $\mathcal{N} = 4$ Super-Yang-Mills theory at strong coupling. By combining spectral data extracted from integrability, with recent advances in computing the AdS Virasoro-Shapiro amplitude, we extract predictions for leading order OPE coefficients on entire Kaluza-Klein (KK-)towers of states. We observe that the appropriately normalised leading order OPE coefficients for all states in a given KK-tower are the same. Furthermore, we also notice that, in many cases, the leading order OPE coefficients of all states in the KK-tower vanish, suggesting a simplification of the physics in this limit.   
}

\maketitle

\section{\label{sec:intro}Introduction}

Integrability and conformal bootstrap have proved to be two of the most efficient tools to obtain perturbative and non-perturbative CFT-data of many CFTs, and in particular, planar 4D $\mathcal{N} = 4$ supersymmetric Yang-Mills theory (SYM)~\cite{Rattazzi:2008pe, El-Showk:2012cjh, Kos:2016ysd, Gromov:2009tv, Gromov:2013pga, Cavaglia:2014exa, Basso:2015zoa, Cavaglia:2018lxi,Jiang:2019zig,Cavaglia:2021mft,Grabner:2020nis, Julius:2021uka, Bercini:2022jxo, Cavaglia:2021eqr,Ekhammar:2021pys}.
Recently, the combination of the two techniques, called bootstrability, has yielded results beyond what is obtainable by either of the techniques alone~\cite{Cavaglia:2021bnz,Cavaglia:2022qpg,Cavaglia:2022yvv,Caron-Huot:2022sdy,Niarchos:2023lot}. 

With these serving as inspiration, in this paper, we perform a careful analysis of the CFT-data of planar 4D $\mathcal{N} = 4$ SYM at strong coupling. First we extract strong coupling spectral information using the integrability-based quantum spectral curve (QSC) method~\cite{Gromov:2013pga, Gromov:2015wca}, in particular its implementation developed in~\cite{Gromov:2023hzc}. We then inject this spectral information into constraints on the CFT-data at strong coupling, obtained in~\cite{Alday:2022uxp,Alday:2022xwz,Alday:2023mvu,Fardelli:2023fyq}.

This allows us to solve for the OPE coefficients, and we obtain leading order expressions at strong coupling, for the OPE coefficients of \emph{entire} Kaluza-Klein (KK-)towers of local operators in $\mathcal{N} = 4$ SYM. 
Our results can be used to extract further constraints on the CFT-data, which could potentially be checked by generalising the computations of~\cite{Fardelli:2023fyq}.
Finally, we observe that leading order OPE coefficients of many \emph{entire} KK-towers vanish, which hits at the possibility of a drastic simplification in the CFT-data of planar 4D $\mathcal{N} = 4$ SYM at strong coupling.

\paragraph{Structure of the paper.} In section \ref{sec:Setup} we describe the four-point function which we consider, outline the current knowledge of the spectrum of $\mathcal{N}=4\  \text{SYM}$ at strong coupling and discuss the constraints on the CFT data at strong coupling. Then in section \ref{sec:Results} we present 
our main results: a method to obtain $R$-charge independent average formulas constraining the CFT-data at strong coupling, and explicit results for leading order OPE coefficients. 
Finally, in section \ref{sec:Discussion} we discuss consequences of our analysis and which potential questions 
that stem from them.

\section{Setup}
\label{sec:Setup}

We will restrict ourselves to planar 4D $\mathcal{N} = 4$ SYM, which is obtained by taking the Yang-Mills coupling $g_\mathtt{YM}\to 0$ and the rank of the gauge group $\mathrm{SU}(N)$, $N\to\infty$ in such a way that a particular combination, called the 't Hooft coupling $\lambda\equiv g_{\mathtt{YM}}^2\,N$ is kept finite. 

Operators in the theory transform under representations of the group $\mathrm{PSU}(2,2|4)$. A state is characterised by the quantum numbers: $[\Delta\;\ell_1\;\ell_2\;q_1\;p\;q_2]$. Here $\Delta$ is the scaling dimension and $\ell_1, \ell_2$ are Lorentz spin labels
of the 4D conformal group $\mathrm{SO}(4,2)$, and $q_1, p, q_2$ are Dynkin labels of $\mathrm{SO}(6)_R$ $R$-symmetry group. 

\subsection{Observable}
We consider four-point functions of four protected operators $\mathcal{O}_{k}$. These are Lorentz scalars which transform in the rank-$k$ symmetric traceless representation of the $\mathrm{SO}(6)_R$ group. Their scaling dimension is $\Delta = k$, and is protected by supersymmetry.

In particular, we consider the four-point function \cite{Dolan:2001tt, Arutyunov:2002fh, Dolan:2003hv, Dolan:2004iy, Caron-Huot:2018kta}
\begin{align}\label{eqn:22kkdef}
\langle \mathcal{O}_2(x_1, y_1) \mathcal{O}_2(x_2, y_2)\mathcal{O}_{k}(x_3, y_3) \mathcal{O}_{k} (x_4, y_4) \rangle = \frac{(y_{12})^2 (y_{34})^{k}} {(x_{12}^2)^2 (x_{34}^2)^{k}} \mathcal{G}_{\{22kk\}}(U, V; \alpha, \bar{\alpha})\;,
\end{align}
where $U \equiv \frac{x_{12}^2 x_{34}^2}{x_{13}^2 x_{24}^2} \equiv z \bar{z}$, $V \equiv \frac{x_{14}^2 x_{23}^2}{x_{13}^2 x_{24}^2} \equiv (1-z)(1-\bar{z})$ and $\frac{y_{12} y_{34}}{y_{13} y_{24}} \equiv \alpha \bar{\alpha}$, $\frac{y_{14} y_{23}}{y_{13} y_{24}} \equiv (1-\alpha)(1-\bar{\alpha})$ are cross-ratios.
Here and below, we follow the conventions of~\cite{Fardelli:2023fyq}.
The four-point function can be further decomposed as~\cite{Dolan:2003hv}
\begin{align}
    \mathcal{G}_{\{22kk\}}(U, V; \alpha, \bar{\alpha}) = \mathcal{G}_{\{22kk\}}^{\mathtt{free}}(U, V; \alpha, \bar{\alpha}) + \frac{(z - \alpha)(z - \bar{\alpha})(\bar{z}-\alpha)(\bar{z}-\bar{\alpha})}{(z\bar{z})^2 (\alpha \bar{\alpha})^2} \mathcal{T}(U, V)\;,
\end{align}
where $\mathcal{G}_{\{22kk\}}^{\mathtt{free}}(U, V; \alpha, \bar{\alpha})$ is the free theory contribution which can be evaluated by performing Wick contractions~\cite{Fardelli:2023fyq}, and $\mathcal{T}(U, V)$ is a reduced correlator which captures the non-trivial contribution to the four-point function. 

Furthermore, we consider separately the contributions of short, or protected supermultiplets and long, or unprotected supermultiplets to the reduced correlator:
\begin{gather}
    \mathcal{T}(U,V) = \mathcal{T}(U,V)^{\mathtt{short}} + \mathcal{T}(U,V)^{\mathtt{long}}\;.
\end{gather}
We are interested in the contribution of unprotected operators, whose scaling dimension depends on the 't Hooft coupling $\lambda$.
The reduced correlator admits an operator product expansion (OPE) in two channels ($s$ and $t$). In the respective channels, we get
\begin{align}\label{eqn:OPE}
\begin{split}
    \mathcal{T}(U, V)_{\{22kk\}}^{\mathtt{long}} &= \sum_{\Delta, \ell} C_{s} G_{\Delta + 4, \ell}^{(0,0)} (z, \bar{z})\;, \\
    \mathcal{T}(U, V)_{\{2k2k\}}^{\mathtt{long}} &= (\alpha \bar{\alpha})^{\frac{2-k}{2}}\sum_{\Delta, \ell} C_{t} G_{\Delta + 4, \ell}^{(k-2,2-k)} (z, \bar{z})
    \;.
\end{split}
\end{align}
where we introduce the conformal blocks~\cite{Dolan:2003hv}
\begin{align}
        G_{\Delta, l}(z, \bar{z})^{(r,s)} &= \frac{z \bar{z}}{\bar{z} - z} \left[\kappa_{\frac{\Delta-l-2}{2}}^{r,s}(z)\kappa_{\frac{\Delta+l}{2}}^{r,s}(\bar{z}) \right. \- \left. \kappa_{\frac{\Delta+l}{2}}^{r,s}(z) \kappa_{\frac{\Delta-l-2}{2}}^{r,s}(\bar{z})  \right] \;, \\
    \kappa_{h}^{r,s}(z) &\equiv z^h {}_{2}F_1 \left(h + \frac{r}{2}, h+\frac{s}{2}; 2h, z \right)\;.
\end{align}
In the expressions~\eqref{eqn:OPE}, the sum runs over the twists $T\equiv \Delta -\ell$, and ``spins'' $\ell$ of the exchanged unprotected operators.   
The spin $\ell$ corresponds to equal Lorentz spin-labels $[\ell\;\ell]$.
In the $s$-channel the exchanged operators have even spin and $R$-symmetry labels $[0\;0\;0]$. In the $t$-channel, exchanged operators can have either odd or even Lorentz spin, and have $R$-symmetry labels $[0\;p\;0]$, where $p \equiv k - 2$.
In the following, we refer to $p$ as the $R$-charge.

Furthermore, we focus only on of the exchange of the single-trace ``stringy'' operators whose anomalous dimension scales as $\Delta \sim \lambda^{1/4}$ at strong coupling. In the following we discuss the conformal data for only such operators.  The OPE coefficients for the ``stringy'' operators $C_s$ and $C_t$ can be parameterised as~\cite{Fardelli:2023fyq}:
\begin{align}\label{eqn:CsCtnorm}
\begin{split}
    C_s &=\frac{\pi^3\,(-1)^{\ell}\, T^{2 {p }+6}\, 2^{-2 \ell-2p-2 T-12}}{(\ell+1)\, \Gamma({p+1})\, \Gamma({p+2})\, \sin ^2\left(\frac{\pi\, T}{2}\right)}\, f_s
    \;,\\
    C_t &= \frac{\pi^3\,(-1)^{\ell}\, T^{2 {p}+6}\, 2^{-2 \ell-2p-2 T- 12}}{(\ell+1)\, \Gamma({p+1})\, \Gamma({p+2})\, \sin ^2\left(\frac{\pi\, (p+2)}{2}+\frac{\pi\, T}{2}\right)}\, f_t
    \;.
\end{split}
\end{align}
Here $T$ denotes the twist of the exchanged operator, and $f_t$ and $f_s$ are reduced OPE coefficients.
The OPE coefficient $C_t$ is the square of three-point structure constants $\langle \mathcal{O}_2\,\mathcal{O}_k\, \mathcal{O}_\Delta \rangle^2 $, while $C_s$ is the product $\langle \mathcal{O}_2\,\mathcal{O}_2\, \mathcal{O}_\Delta \rangle\times\langle \mathcal{O}_k\,\mathcal{O}_k\, \mathcal{O}_\Delta \rangle$.
Here $\mathcal{O}_\Delta$ refers to the exchanged operator.
As such, $C_t$ can be subjected to positivity constraints.
The reduced OPE coefficients $f$ have the strong coupling expansions
\begin{align}\label{eqn:fdef}
    f_\mathtt{chan} = f_{\mathtt{chan};0} + \frac{f_{\mathtt{chan};1}}{\lambda^{1/4}} + \frac{f_{\mathtt{chan};2}}{\lambda^{1/2}}+ \dots \;.
    \;,
\end{align}
where $\mathtt{chan}\in\{s,t\}$.
Moreover, $f_{{t;0}}$ becomes equal to $f_{{s;0}}$, when $k = 2 \Rightarrow p=0$~\cite{Fardelli:2023fyq}. 
Therefore,
for simplicity of notation, we use a single notation $f_0$, and understand it to mean $f_{t;0}$ when $p\neq 0$ and $f_{s;0} = f_{t;0}$ when $p=0$. 
These reduced OPE coefficients are the main observables targeted by our calculation.

\subsection{Spectrum at strong coupling}\label{sec:spec}
Let us consider in detail the dimensions of unprotected operators exchanged in the OPE~\eqref{eqn:OPE}, at strong coupling, in the planar limit. 

We focus on the single-trace operators which are also called the ``stringy'' operators. At strong coupling, the dimensions these operators have the expansion
\begin{align}\label{eqn:DelStrong}
    \Delta = 2\,\sqrt{\delta} \lambda^{1/4} + d_0 + \frac{d_1}{\sqrt{\delta}\, \lambda^{1/4}} + \dots\;.
\end{align}
where $\delta$, a positive integer, is the string mass level \cite{Gubser:1998bc}. 
Thus, along with the spin $\ell$ and the $R$-charge $p$, it is a good label for classifying the states exchanged in the OPE~\eqref{eqn:OPE}. In general, however, there could be multiple states, which have the same labels $(\delta\;\ell\;p)$. In order to break this degeneracy, we need to look at higher orders in the strong coupling expansion~\eqref{eqn:DelStrong} of $\Delta$.

In~\cite{Gromov:2023hzc}, the dimensions of the lowest lying 219 states in $\mathcal{N} = 4$ SYM were computed numerically using the QSC~\cite{Gromov:2013pga,Gromov:2014caa,Gromov:2015wca,Hegedus:2016eop}. The numerical spectrum was fitted at strong coupling, and expressions for $d_0$, and $d_1$ in the expansion~\eqref{eqn:DelStrong} were obtained for many of these operators. It was argued in~\cite{Gromov:2023hzc} that value of $d_0 = -2$ is universal for all states in planar $\mathcal{N} = 4$ SYM. Therefore, to lift the degeneracies, one must know the value of $d_1$, for every state with the same labels $(\delta\;\ell\;p)$.

While the precise value of $d_1$ for the various degenerate operators may not be known, it is nevertheless possible to enumerate the degeneracies of operators with the same labels $(\delta\;\ell\;p)$. This was done in~\cite{Alday:2023flc}
by mapping the four-point function $\langle \mathcal{O}_{k_1}\,\mathcal{O}_{k_2}\,\mathcal{O}_{k_3}\,\mathcal{O}_{k_4}\rangle$ in planar 4D $\mathcal{N} = 4$ at strong coupling, to a four-point closed string amplitude in 
AdS${}_5\times S^5$  (the AdS Virasoro-Shapiro amplitude), in the flat-space limit. 
Then the stringy operators exchanged in the OPE~\eqref{eqn:OPE} map to massive string states in flat-space, compactified on $S^5$.
To count these states, firstly representations of $\mathrm{SO}(9)$, the massive little group for $\mathbb{R}^{1,9}$, were enumerated. 
Of these, only representations of $\mathrm{SO}(4) \times \mathrm{SO}(5)$, 
were considered, corresponding to the split into $AdS_5$ and $S^5$.
Lastly, compactification of five-directions into into $S^5$ replaced every representation of $\mathrm{SO}(5)$ with Dynkin labels $[m\;n]$, by a KK-tower of $\mathrm{SO}(6)$ representations~\cite{Bianchi:2003wx}:
\begin{multline}
\label{eqn:KKtowers}
    \mathrm{KK}_{[m\;n]}
    =\sum_{r=0}^m \sum_{s=0}^n \sum_{p=m-r}^{\infty}\left[\ell_1, \ell_2, r+n-s, p, r+s\right] \\
    +\sum_{r=0}^{m-1} \sum_{s=0}^{n-1} \sum_{p=m-r-1}^{\infty}\left[\ell_1, \ell_2, r+n-s, p, r+s+1\right]\;.
\end{multline}
Thus, in~\cite{Alday:2023flc}, a counting function~$\mathtt{count}(\delta,\ell)$ was obtained, which,
for every value of $\delta$ and $\ell$, counts the number of KK-towers $\mathrm{KK}_{[m\;n]}$. 

Out of all the choices for $\mathrm{SO}(5)$ labels, notice that only $\mathrm{KK}_{[m\;0]}$ contains KK-towers of states whose $R$-symmetry labels are of form $[0\;p\;0]$.
Therefore, we are only interested in such KK-towers. 
Furthermore, since $\mathrm{KK}_{[m\;0]} = \sum_{p = m}^\infty [0\;p\;0] + \cdots$, for states in such a KK-tower, the $R$-charge $p\geq m$. 

Let $N_{m^\prime}$ be the number of KK-towers of type $\mathrm{KK}_{[m\;0]}$, with $0\leq m\leq m^{\prime}$, outputted by $\mathtt{count}(\delta,\ell)$.
It follows that the degeneracy of states with labels $(\delta,\ell,p)$ is $N_p$. 
Let $M$ be the maximal value of $m$, for which $\mathtt{count}(\delta,\ell)$ outputs a KK-tower of type $\mathrm{KK}_{[m\;0]}$. 
Since there can be no new KK-towers for higher values of the $R$-charge, for all $p \geq M$, and the degeneracy states with labels $(\delta\;\ell\;p)$ is $N_{M}$.
Thus, in general, for a given $\delta$ and $\ell$ the degeneracy of KK-towers $N_m$ increases from $p = 0$ till $p=M$, from when it is a constant $N_M$, for all higher $p$.

In~\cite{Gromov:2023hzc}, the counting of~\cite{Alday:2023flc} was confirmed by explicit computation of scaling dimensions of operators in $\mathcal{N} = 4$ SYM; it was shown that the dimensions of states at strong coupling indeed organised themselves into exactly the same KK-towers as predicted in~\cite{Alday:2023flc}.
Furthermore, it was observed in~\cite{Gromov:2023hzc} that the quadratic Casimir of $\mathrm{PSU}(2,2|4)$ (given in equation~\eqref{eqn:Casimir}),
was a good classifier of states into KK towers.
Denote the strong coupling expansion of the quadratic Casimir as
\begin{align}
    J^2 = 2\,\delta \sqrt{\lambda} + j_1 + \frac{j_2}{\sqrt{\lambda}}+ \dots\;.
\end{align}
In particular, the constant term in the strong coupling expansion of the Casimir $j_1$, was observed to be the same for every state in a KK-tower. This led to a conjecture, predicting the values of $d_1$ for all states in a KK-tower, given the value of $j_1$ on the tower. We have~\cite{Gromov:2023hzc}
\begin{multline}\label{eqn:CasConj}
	d_1 = \frac{p^2}{4}+\frac{p}{4}\left(q_1+q_2+4\right) +\frac{1}{16}\big[16  -2 \ell_1\left(\ell_1+2\right)-2 \ell_2\left(\ell_2+2\right) \\+3 q_1\left(q_1+4\right)+3 q_2\left(q_2+4\right)+2 q_1 q_2\big]+\frac{j_1}{2}\;.
\end{multline}

\subsection{Constraints at strong coupling}

By exploiting the supersymmetry, conformal symmetry and Regge behaviour of the correlator~\eqref{eqn:22kkdef}, the flat-space limit and number theoretic properties of the corresponding string amplitude (the AdS Virasoro-Shapiro amplitude), in~\cite{Alday:2022uxp,Alday:2022xwz,Alday:2023flc,Alday:2023jdk,Alday:2023mvu,Fardelli:2023fyq}, constraints on the various combinations of the strong coupling expansion coefficients of the OPE coefficients~\eqref{eqn:fdef} and the scaling dimensions~\eqref{eqn:DelStrong}, of stringy operators, were obtained. 

These ``average'' formulas, constrain the sum over the particular combinations of expansion coefficients of the CFT-data, of all states with the same labels $(\delta\;\ell\;p)$. Moreover, analytical formulas were found for operators on the same odd- or even-spin Regge trajectories. The string mass level $\delta$ and spin $\ell$ of states on the same even- or odd-spin Regge trajectory are related as
\begin{align}
    \ell = 2\,(\delta - n)\;, \quad\text{or}  \quad \ell = 2\,(\delta - n) - 1\;,
\end{align}
where $n$ is a positive integer called the even- or odd-spin Regge trajectory number respectively.

Our aim is to extract the strong coupling expansion coefficients of OPE coefficients~\eqref{eqn:fdef} of particular states. 
On the first even-spin Regge trajectory, \textit{i.e.}, the trajectory with $\ell = 2\,(\delta -1)$, the degeneracy of states is unity~\cite{Alday:2023flc}. Therefore, the average formulas directly give us the required information about the strong coupling expansion coefficients. However, on higher even- and odd-spin Regge trajectories, this is not the case. Thus, we need to ``unmix'' the average formulas, \textit{i.e.} extract the CFT-data of individual states that enter the average formulas. In the next section, we achieve precisely this, by injecting the spectral information extracted from~\cite{Gromov:2023hzc} into the average formulas, thereby unmixing them, and extracting predictions for strong coupling expansion coefficients of the OPE coefficients.

\section{Results}
\label{sec:Results}
In this section, we present our main results. 
Firstly, for states with labels $(\delta\;\ell\;p)$, we obtain average formulas of the type $\langle f_0\;j_1 \rangle$ and $\langle f_0\;j_1^2 \rangle$, 
where $j_1$ is the sub-leading Casimir, and 
$f_0$ is the leading order reduced OPE coefficient.
Then, injecting information on $j_1$, extracted from~\cite{Gromov:2023hzc} into these formulas, we extract predictions for $f_0$ that hold for \emph{entire} KK-towers of states.  

Our focus will be the leading order reduced OPE coefficients $f_0$. Therefore, only even-spin Regge trajectories are relevant, as $f_0$ vanishes on odd-spin Regge trajectories~\cite{Fardelli:2023fyq}.
As the OPE coefficients on the leading even-spin Regge trajectory, \textit{i.e.}~with $\ell = 2\,(\delta -1)$, are directly known from the average formulas of~\cite{Alday:2022uxp,Alday:2022xwz,Alday:2023mvu,Fardelli:2023fyq}, our focus will be on states in the second and third even-spin Regge trajectories, \textit{i.e.}~those with $\ell = 2\,(\delta -2)$ and $\ell = 2\,(\delta - 3)$ respectively.  

\subsection{Average formulas involving the sub-leading Casimir}

Consider a formula for $\langle f_0\,d_1\rangle$ for states with labels $(\delta\;\ell\;p)$, on the $n^\text{th}$ even-spin Regge trajectory. In general we have 
\begin{align}\label{eqn:f0d1sum}
    \langle f_0\,d_1\rangle_{\ell = 2(\delta - n)} \equiv \sum_{I=1}^{N_p} f_{0}^{I}\,d_{1}^{I}
    = g_{n}(\delta,p)
    \;,
\end{align}
where $f_{0}^{I}$ and $d_{1}^{I}$ are the respective values of $f_0$ and $d_1$ for the $I^\text{th}$ state in the sum over $N_p$ degenerate states, and $g_{n}$ is a  function of $\delta$ and $p$.
Expressions for $g_{n}(\delta, 0)$ are known on the first seven even-spin Regge trajectories~\cite{Alday:2022xwz}, and those for $g_{n}(\delta,p)$ are known on the first even-spin Regge trajectory~\cite{Fardelli:2023fyq}.
It was shown in~\cite{Alday:2023flc, Fardelli:2023fyq}, that the individual $f_{0}^{I}$ cannot depend on $p$. 
The $p$-dependence of $d_{1}^{I}$ is fixed by the conjecture of~\cite{Gromov:2023hzc}, given in equation~\eqref{eqn:CasConj} as
\begin{align}\label{eqn:d1Regge}
    d_{1}^{I} = \frac{p^2}{4} +p -\delta^2 + \delta\,(2\, n-1) -n^2 + n + 1 + \frac{j_{1}^{I}}{2}\;.
\end{align}
Notice in particular that the $p$-dependence is the same for all states with a given $\delta$ and $n$.
Plugging this into equation~\eqref{eqn:f0d1sum}, we get
\begin{align}
    \langle f_{0}\,j_{1}\rangle_{\ell = 2(\delta - n)}
    = 2\,g_{n}(\delta,p) - 
    2\,\langle f_0\rangle_{\ell = 2\,(\delta - n)}
    \bigg[\frac{p^2}{4} +p 
    -\delta^2 + \delta\,(2 n-1) -n^2 + n + 1\bigg]
    \;.
\end{align}
Since both $f_0$ and $j_1$, and therefore the LHS is $p$-independent, the RHS must be $p$-independent too. This means that the $p$-dependence of $g_n(\delta,n,p)$ should be 
\begin{align}
    g_n(\delta,p) = g_n(\delta,0) + \langle f_0 \rangle \bigg[\frac{p^2}{4} + p\bigg]\;,
\end{align}
so as to cancel the $p$-dependence of the other terms in the RHS.
Therefore, we get
\begin{align}\label{eqn:f0j1gen}
    \langle f_{0}\,j_{1}\rangle_{\ell = 2(\delta - n)}
    = 2\,g_{n}(\delta,0) + 
    2\,\langle f_0\rangle_{\ell = 2\,(\delta - n)}
    \bigg[
    \delta^2 - \delta\,(2\,n-1) + n^2 - n - 1\bigg]
    \;.
\end{align}
Thus, starting from expressions $\langle f_0\,d_1 \rangle$, for states with $p=0$, given in~\cite{Alday:2022xwz}, one can extract a formula for $\langle f_0\,j_1 \rangle$, valid for all states that share the same value of $j_1$. In particular such a formula will be valid, even for states with $p\neq 0$. 
For the first three even-spin Regge trajectories, we display the explicit expressions for $\langle f_0\,j_1\rangle$ in equations~\eqref{eqn:f0j1t1} --~\eqref{eqn:f0j1t3}.

We can reverse the logic now, to obtain expressions for $\langle f_0\,d_1\rangle$, for states with $p \neq 0$. 
Starting with equation~\eqref{eqn:f0j1gen}, we can plug in $j_1$ in terms of $d_1$ from~\eqref{eqn:d1Regge}, assuming this time, that $p\neq 0$, to get
\begin{align}
    \langle f_0\, d_{1}\rangle_{\ell = 2\,(\delta - n)}
    = g_{n}(\delta,0) +\langle f_0\rangle_{\ell = 2\,(\delta - n)}\,
    \bigg[
     \frac{p^2}{4} + \,p
    \bigg]
    \;.
\end{align}
For the first even-spin Regge trajectory, it can be checked that this formula gives us exactly what was obtained in~\cite{Fardelli:2023fyq}.

One can make the same arguments as above to the formulas for $\langle f_0\,d_1^2\rangle$ for states with $p=0$ that can be extracted from~\cite{Alday:2023mvu}.
First, one can extract $p$-independent expressions for $\langle f_0\,j_1^2\rangle$.
Such formulas, for the first three even-spin Regge trajectories are provided in equations~\eqref{eqn:f0j1sqt1} --~\eqref{eqn:f0j1sqt3}.
Then one can reverse the logic to extract formulas of the type $\langle f_0\,d_1^2\rangle$, for states with $p\neq 0$.
We provide such formulas for the first three even-spin Regge trajectories in equations~\eqref{eqn:f0d1sqt1} --~\eqref{eqn:f0d1sqt3}. 
These formulas are a prediction which, in principle, could be checked by extending the methods of~\cite{Fardelli:2023fyq} to the next order.

Let us explore the consequences of the above exposition.
We have at our disposal, $p$-independent average formulas involving sums of combinations of $f_0$ and $j_1$ over degenerate states.
Consider such an average formula of the form $\langle f_0\,j_1^\alpha\rangle$, where $\alpha$ is a integer $>0$.
For every choice of labels $(\delta,\ell,p)$, the LHS of this formula involves the sum over $N_p$ degenerate states.
Due to $p$-independence, the RHS of this formula remains the \emph{same} for all choices of $p$. We have
\begin{align}\label{eqn:avgformula}
    \langle f_0 \,j_1^\alpha \rangle \equiv \sum_{I = 1}^{N_p}  f_0^I\,\left(j_1^I\right)^\alpha = c_n(\delta)\;.
\end{align}
Here $c_n(\delta)$ is a constant, depending on spin $\ell$ through the Regge trajectory number $n$ and the string mass level $\delta$.
The average formula~\eqref{eqn:avgformula}, involves $j_1$, which has the \emph{same} value for all states in a KK-tower~\cite{Gromov:2023hzc}.
This suggests to consider KK-towers of states that share the same $\delta$ and $\ell$, rather than the individual states that live in these KK-towers. 
As elucidated in section~\ref{sec:spec}, for every choice of $\delta$ and $\ell$, there exists a non-negative integer $M$, so that the degeneracy of KK-towers 
is a constant $N_M$ for all $p\geq M$.
Thus, for all $p\geq M$, the LHS of this average formula involves a sum over $M$ degenerate KK-towers. 
Therefore, for all $p\geq M$, the formula~\eqref{eqn:avgformula} becomes
\begin{align}\label{eqn:avgformulaM}
    \langle f_0 \,j_1^\alpha \rangle
     \equiv \sum_{I = 1}^{N_M} f_0^I\,\left(j_1^I\right)^\alpha = c_n(\delta)\;.
\end{align}
Suppose we have a system of $N_M$ such equations, with $\alpha = 0,\dots,N_M-1$.
The values of $f_0$ that solve this system will hold for \emph{all} states in the corresponding KK-towers, with $R$-charge $p\geq M$.
For each value of the $R$-charge $p < M$, since the number of KK-towers is $N_p < N_M$, we would need a smaller system of at least $N_p$ equations of the form~\eqref{eqn:avgformula}. Therefore, for KK-towers that contain states with $p<M$, it is not guaranteed that $f_0$ has the same value as those states with $R$-charge $p\geq M$.

\subsection{Predictions on KK-towers and vanishing OPE coefficients}

Armed with the average formulas obtained using the methods of the previous section, we now proceed to extract predictions for the leading order reduced OPE coefficients $f_0$. 

The relevant KK-towers on the first three even-spin Regge trajectories, are summaried in table~\ref{tab:KK2} below, we have
\begin{xltabular}{\textwidth}{c|C|C|C}
    \diagbox{$\delta$}{$\ell$} & 0 & 2 & 4 \\
    \midrule\midrule
     1  &   \mathrm{KK}_{[0\;0]} &  &    \\[0.2em]
     2  &   2\,\mathrm{KK}_{[0\;0]} + \mathrm{KK}_{[2\;0]} &   \mathrm{KK}_{[0\;0]} &  \\[0.2em]
     3  &   6\,\mathrm{KK}_{[0\;0]} + 2\,\mathrm{KK}_{[1\;0]} + 4\,\mathrm{KK}_{[2\;0]} + \mathrm{KK}_{[4\;0]} &   4\,\mathrm{KK}_{[0\;0]} + 3\,\mathrm{KK}_{[2\;0]} &   \mathrm{KK}_{[0\;0]}  \\[0.2em]
    \caption{Number of KK-towers of the type $\mathrm{KK}_{[m,0]}$ for different values of $\delta$ and $\ell$ obtained by evaluating $\mathtt{count}(\delta,\ell)$ of \cite{Alday:2023flc}.
    \label{tab:KK2}
    }
\end{xltabular}
In the sequel, we will consider three cases: $\delta= 2,\,\ell = 0$ and $\delta = 3,\,\ell = 2$ on the second even-spin Regge trajectory, and $\delta = 3,\,\ell = 0$ on the third even-spin Regge trajectory.

\paragraph{KK-towers with $\delta = 2$ and $\ell = 0$.}
From the counting function $\mathtt{count}(\delta,\ell)$ of~\cite{Alday:2023flc}, we get the following relevant KK-towers:
\begin{align}
    \mathtt{count}(2,0) = 2\,\mathrm{KK}_{[0\;0]} + \mathrm{KK}_{[2\;0]} + \cdots
    \;.
\end{align}
Here the ellipsis denotes KK-towers not of the form $\mathrm{KK}_{[m\;0]}$, \textit{i.e.} those which do not contain states with $R$-symmetry labels of the form $[0\;p\;0]$.
Thus, for $p\geq 2$, there are 3 states which enter the average formulas for $\langle f_0 \rangle$~\eqref{eqn:f0t2} of~\cite{Alday:2022uxp}, for $\langle f_0\,j_1\rangle$~\eqref{eqn:f0j1t2} and for $\langle f_0\,j_1^2\rangle$~\eqref{eqn:f0j1sqt2}.
The values of $j_1$ for these three towers of states
can be extracted from~\cite{Gromov:2023hzc}. 
They are given in table~\ref{tab:EvenReggeStates}, and are repeated below:
\begin{align}\label{eqn:j1del2ell0}
    j_1^{[0\;0]_1} = 2\;,\quad j_1^{[0\;0]_2} = 14\;,\quad j_1^{[2\;0]} = 2
    \;.
\end{align}
In the above expression, we denote the particular KK-tower corresponding to the value of $j_1$ in the superscript. Where there is a multiplicity, we have included an extra multiplicity label.
Plugging the expressions from~\eqref{eqn:j1del2ell0} into the average formulas for $\langle f_0\rangle$ and $\langle f_0\,j_1\rangle$, we get the solution
\begin{align}\label{eqn:del0ell2preres}
    f_{0}^{[0\;0]_1} = - f_{0}^{[2\;0]}\;, \quad f_{0}^{[0\;0]_2} = \frac{1}{4}\;.
\end{align}
Requiring that the leading order OPE coefficients are $\geq 0$ immediately sets
\begin{align}\label{eqn:del0ell2res}
    f_{0}^{[0\;0]_1} = f_{0}^{[2\;0]} = 0\;, \quad f_{0}^{[0\;0]_2} = \frac{1}{4}\;.
\end{align}
This result is consistent with our formula~\eqref{eqn:f0j1sqt2} for $\langle f_0\,j_1^2\rangle$, and thus serves as a check for it.
To clarify the notation, when we use $f_{0}^{[m\;0]}$ we are only referring to the KK-tower $\sum_{p = m}^\infty [0\;p\;0]$, and not the other KK-towers that are obtained when evaluating $\mathrm{KK}_{[m\;0]}$ using~\eqref{eqn:KKtowers}.
The value of $j_{1}^{[m\;0]}$ however, is valid for \emph{all} KK-towers obtained when evaluating $\mathrm{KK}_{[m\;0]}$ using~\eqref{eqn:KKtowers}, and therefore, this notation may be used for all of them.

Strictly speaking, the above solution is valid only when $p\geq 2$, as explained in the previous section.
For the case $0\leq p<2$, one has only 2 states coming from the 2 $\mathrm{KK}_{[0\;0]}$, and therefore should repeat the above procedure with only 2 variables $f_0$. Doing so gives the same result, and therefore~\eqref{eqn:del0ell2res} gives leading order predictions for OPE coefficients on 3 \emph{entire} KK-towers.
For the $p = 0$ case, predictions for $f_0$ were obtained in~\cite{Gromov:2023hzc,Alday:2023mvu}, and our results are consistent with them. 

Notice that 2 out of 3 entire KK-towers of leading order OPE coefficients vanish.
It would be interesting to note whether they vanish to higher orders as well. 

\paragraph*{KK-towers with $\delta = 3$ and $\ell = 2$.} 
From the counting function of~\cite{Alday:2023flc}, we get
\begin{align}
    \mathtt{count}(3,2) = 4\,\mathrm{KK}_{[0\;0]} + 3\,\mathrm{KK}_{[2\;0]} + \cdots
    \;.
\end{align}
Thus, generically, we have 7 KK-towers of states. 
The values of $j_1$, extracted from~\cite{Gromov:2023hzc}, given in table~\ref{tab:EvenReggeStates}, are displayed below. We have
\begin{align}
    j_1^{[0\;0]_1} = j_1^{[2\;0]_1} = j_1^{[2\;0]_2} = j_1^{[2\;0]_3} = 18 \;, \quad
    j_1^{[0\;0]_2} = 36 \;, \quad
    j_1^{[0\;0]_3} = j_1^{[0\;0]_4} =  28
    \;.
\end{align}
Notice furthermore, that corresponding states in the KK-towers $[0\;0]_3$ and $[0\;0]_4$, with the same $R$-charge, are \emph{exactly degenerate}, \textit{i.e} their scaling dimensions are the same, non-perturbatively, and in particular to all orders in perturbation theory. The  corresponding  states in the KK-towers $[2\;0]_2$ and $[2\;0]_3$ are also exactly degenerate.
The exact degeneracy is due to a symmetry of the underlying integrability structure~\cite{Marboe:2017dmb, Marboe:2018ugv} (see also~\cite{Gromov:2023hzc}).

The OPE coefficients of exactly degenerate states will also be the same.
Thus, $f_0^{[0\;0]_3}=f_0^{[0\;0]_4}$ and $f_0^{[2\;0]_2}=f_0^{[2\;0]_3}$.
Plugging the values of $j_1$ from above into equations~\eqref{eqn:f0t2},~\eqref{eqn:f0j1t2} and~\eqref{eqn:f0j1sqt2},
we get
\begin{align}
    f_0^{[0\;0]_1} = - f_0^{[2\;0]_1} - 2 f_0^{[2\;0]_2}\;, \quad
    f_0^{[0\;0]_2} = \frac{243}{1024} \;, \quad
    f_0^{[0\;0]_3} = f_0^{[0\;0]_4} = \frac{135}{1024}
    \;.
\end{align}

Next, imposing positivity of the OPE coefficient, immediately gives
\begin{align}
\label{eqn:f0del3ell2}
    f_0^{[0\;0]_1} = f_0^{[2\;0]_1} = f_0^{[2\;0]_2} = f_0^{[2\;0]_3} = 0
    \;.
\end{align}
Again, strictly speaking the above result holds only when $p\geq 2$. However, it turns out that the same holds when $0\leq p <2$ as well, when there are a lesser number of KK-towers entering the average formulas.
Furthermore, 4 out of 7 entire KK-towers of leading order OPE coefficients vanish. 
For the states with $p = 0$, our results are consistent with, and interestingly, they saturate the bounds on the leading order OPE coefficients obtained in~\cite{Gromov:2023hzc}.

\paragraph{KK-towers with $\delta = 3$ and $\ell = 0$.}

Again here, we start with the counting function of~\cite{Alday:2023flc,AHSPrivate}, which gives
\begin{align}
    \mathtt{count}(3,0) = 6\,\mathrm{KK}_{[0\;0]} + 2\,\mathrm{KK}_{[1\;0]} + 4\,\mathrm{KK}_{[2\;0]} + \mathrm{KK}_{[4\;0]} + \cdots
    \;.
\end{align}
Thus, in general, there can be up to 13 KK-towers of states. 
We display the values of $j_1$ for these states, extracted from~\cite{Gromov:2023hzc}, given in table~\ref{tab:EvenReggeStates}, below:
\begin{align}
\begin{split}
    j_1^{[0\;0]_1} &= j_1^{[0\;0]_6} = \frac{27}{2} \;, \quad 
    j_1^{[0\;0]_2} = j_1^{[1\;0]_1} = j_1^{[2\;0]_1} = j_1^{[4\;0]} = 0\;, \\
    j_1^{[0\;0]_3} &= j_1^{[1\;0]_2} = j_1^{[2\;0]_2} = 18\;, \quad j_{1}^{[0\;0]_4} = 20\;, \quad
    j_{1}^{[0\;0]_5} = 36\;,\quad j_{1}^{[2\;0]_3} = j_{1}^{[2\;0]_4} = 10\;.
\end{split}
\end{align}
In this case as well, there are some exactly degenerate KK-towers, namely $[0\;0]_1$ and $[0\;0]_6$ are exactly degenerate, and so are ${[2\;0]_3}$ and $[2\;0]_4$. Consequently, we get $f_0^{[0\;0]_1} = f_0^{[0\;0]_6}$ and $f_0^{[2\;0]_3} = f_0^{[2\;0]_4}$.
Since, we are on the third even-spin Regge trajectory, in addition to~\eqref{eqn:f0t3}, of~\cite{Alday:2022uxp}, we also need equations~\eqref{eqn:f0j1t3} and~\eqref{eqn:f0j1sqt3}. Solving, we get
\begin{align}
    \begin{split}
        f_{0}^{[0\;0]_3} &= 
        -\frac{65 }{8}f_0^{[0\;0]_1}-20 f_0^{[0\;0]_2}-20 f_0^{[1\;0]_1}-f_0^{[1\;0]_2}-20 f_0^{[2\;0]_1}-f_0^{[2\;0]_2}-\frac{130 }{9}f_0^{[2\;0]_3}-20 f_0^{[4\;0]}
        \;,\\
        f_{0}^{[0\;0]_4} &= 
        \frac{405 }{64}f_0^{[0\;0]_1}+\frac{81 }{4}f_0^{[0\;0]_2}+\frac{81 }{4}f_0^{[1\;0]_1}+\frac{81 }{4}f_0^{[2\;0]_1}+13 f_0^{[2\;0]_3}+\frac{81 }{4}f_0^{[4\;0]}+\frac{25}{1024}
        \;, \\
        f_{0}^{[0\;0]_5} &= 
        -\frac{13 }{64}f_0^{[0\;0]_1}-\frac{5 }{4}f_0^{[0\;0]_2}-\frac{5 }{4}f_0^{[1\;0]_1}-\frac{5 }{4}f_0^{[2\;0]_1}-\frac{5 }{9}f_0^{[2\;0]_3}-\frac{5 }{4}f_0^{[4\;0]}+\frac{81}{1024}
        \;.
    \end{split}
\end{align}
Requiring that all leading order OPE coefficients are $\geq 0$ immediately gives
\begin{align}\label{eqn:f0del3ell0}
    \begin{split}
        f_{0}^{[0\;0]_1} &= f_{0}^{[0\;0]_2} = f_{0}^{[0\;0]_3} = f_{0}^{[0\;0]_6} = f_{0}^{[1\;0]_1} = f_{0}^{[1\;0]_2}
        = f_{0}^{[2\;0]_1} = f_{0}^{[2\;0]_2} = f_{0}^{[2\;0]_3} \\ = f_{0}^{[2\;0]_4} &= f_{0}^{[4\;0]} = 0 \;,\quad
        f_{0}^{[0\;0]_4} = \frac{25}{1024} \;, \quad
        f_{0}^{[0\;0]_5} = \frac{81}{1024} \;.
    \end{split}
\end{align}
In this case, 11 out of 13 KK-towers of leading order OPE coefficients vanish.
Again, strictly speaking the above expression holds only when $p\geq 4$; in the case of $p = 2$, there will be 12 KK-towers, for $p = 1$, there will be 8 KK-towers, and for $p = 0$, there will be 6 KK-towers. However, it can be checked that the same results hold for all values of the $R$-charge $p$.
For states with $p = 0$, our results are consistent with, and saturate the bounds obtained in~\cite{Gromov:2023hzc}.

\paragraph{In conclusion,} we see that in all the three cases considered, our predictions for the leading order reduced OPE coefficients are a constant on an \emph{entire} KK-tower.
Furthermore, many \emph{entire} KK-towers leading order OPE coefficients vanish: for $\delta = 2,\,\ell = 0$, 2 out 3 towers vanish~\eqref{eqn:del0ell2res}, for $\delta = 3,\,\ell = 2$, 4 out 7 towers vanish~\eqref{eqn:f0del3ell2}, and for $\delta = 3,\,\ell = 0$, 11 out 13 towers vanish~\eqref{eqn:f0del3ell0}. 
It is also noteworthy that the degeneracy of KK-towers with the same value of $j_1$ associated with vanishing OPE coefficients changes as we go from $p = 0$ to higher values of the $R$-charge. This ensures that the leading order reduced OPE coefficient remains constant on the \emph{entire} KK-tower.

\section{Discussion}
\label{sec:Discussion}

In this paper, we have performed a careful analysis of the CFT-data of planar 4D $\mathcal{N} = 4$ SYM at strong coupling, which revealed the following salient points.

Firstly, for all the examples that we studied, the leading order reduced OPE coefficients of \emph{all} states in a KK-tower is the same.
The degeneracy of KK-towers for a given value of $\delta$ and $\ell$, in general, changes with the $R$-charge $p$, up to a certain value $p = M$. 
Above this value, the number of degenerate KK-towers is a constant.
That the leading order reduced OPE coefficient $f_0$ is a constant on a KK-tower, for all states with $R$-charge $p \geq M$ is a consequence of our results. 
However, in all the 23 KK-towers that we studied, we observed that the leading order reduced OPE coefficient $f_0$, remains the same on a KK-tower, for all values of $p<M$ as well.
This result may be expected from a flat-space limit point of view, since the dual string amplitude, in this limit, should not be able to see the effects of compactification, and thus, sees the entire KK-tower, as one state, with one OPE coefficient.

Secondly, we observed that the leading order reduced OPE coefficients $f_0$ of 17 out of 23 \emph{entire} KK-towers vanish. 
From a technical point of view, the positivity of $f_0$ was a very powerful in obtaining this conclusion, as in all three cases that we considered, it was responsible for causing many coefficients $f_0$ to vanish.

Whilst the vanishing of a significant proportion of the leading order reduced OPE coefficients $f_0$ signals a simplification of the CFT-data of planar 4D $\mathcal{N} = 4$ SYM at strong coupling,
one should be careful while interpreting it this way.
This is because the reduced OPE coefficients $f_s$ and $f_t$ need to be multiplied by appropriate normalisation factors in order to get the full OPE coefficient $C_s$ and $C_t$ respectively. 
These normalisation factors are complicated functions of $\lambda$. 
To illustrate, consider the $p=0$ case. In this case, let $\mathcal{C}^2 = C_s = C_t$. At large $\lambda$, we see, from equation~\eqref{eqn:CsCtnorm}, that~\cite{Alday:2022uxp}
\begin{align}\label{eqn:CsqNorm}
    \mathcal{C}^2 \sim \frac{2^{-4\, \sqrt\delta\, \lambda^{{1}/{4}}} \lambda^{{3}/{2}}}{\sin ^2\left({\pi\, \sqrt\delta\, \lambda^{{1}/{4}}}\right)}
    \;.
\end{align}
There are two things to notice here. Firstly, we see that $\mathcal{C}^2$ is exponentially damped and, at large enough $\lambda$, the damping factor is the \emph{same} for \emph{all} states with the same string mass level $\delta$.
Secondly, $\mathcal{C}^2$ contains double poles whenever $\sqrt\delta\,\lambda^{1/4} = n$, where $n$ is an integer. This can be interpreted as being due to the mixing of stringy operators and double trace operators with $n>>1$~\cite{Alday:2022uxp}.
Again, at large enough $\lambda$, the location of the double poles is the \emph{same} for \emph{all} states with the same string mass level $\delta$. 
In general, as we can see from equation~\eqref{eqn:CsCtnorm}, the normalisation factors are completely determined by the twists $T$, which is only one half of the CFT-data. 
Important dynamical information unique to the OPE coefficients is contained in the strong coupling expansion coefficients of the reduced OPE coefficient $f \equiv f_t = f_s$, which multiplies the normalisation factor of $\mathcal{C}^2$ from equation~\eqref{eqn:CsqNorm}.
When $\lambda$ is in the neighbourhood of the double pole of the sine squared function, the exponential damping competes with the singularity, and thus the strong coupling expansion coefficients of $f$ become visible, giving access to the fine structure of the OPE coefficients.\footnote{We thank Dileep Jatkar for discussions related to the above passage.}
In these regions it is possible to see that a significant proportion of OPE coefficients are \emph{subleading} with respect to a minority of them.
It would be very interesting to understand what is the physics behind this.

The leading reduced OPE coefficients $f_0$ are related to the strong coupling expansion of the flat-space Virasoro-Shapiro amplitude in Mellin space~\cite{Costa:2012cb,Goncalves:2014ffa,Alday:2022uxp,Fardelli:2023fyq}.
An example of such a relation for the $p = 0$ case is~\cite{Alday:2022uxp}
\begin{align}
    \alpha_{q, 0}=
    \frac{\Gamma{(6 + 2 \,q})}{8^q}\sum_{\delta=1}^{\infty} \sum_{\ell=0,2}^{2 \delta-2} \frac{ \langle f_0 \rangle }{\delta^{3+2\, q}}
    \;.
\end{align}
Here $\alpha_{q,0}$ is the coefficient of $(s^2 + t^2 + u^2)^q/\lambda^{3/2 + q}$ in the expansion of the flat-space Virasoro-Shapiro amplitude, where $s$, $t$ and $u$ are Mellin-Mandelstam variables. 
Why many $f_0$ in the above sum vanish could be explained by, for instance, some emergent symmetry at strong coupling, due to the flat-space limit of the dual string amplitude. 
If so, can this symmetry be used to predict/count the number of vanishing leading order reduced OPE coefficients?
It appears that the number of vanishing KK-towers is increasing drastically with $\delta$. It would be instructive to compare this rate of growth with the rate of growth of degeneracies of states obtained in~\cite{Alday:2023flc}.

Another important question this raises is whether these reduced OPE coefficients vanish at higher orders in perturbation theory as well. 
In particular, it was shown in~\cite{Alday:2023mvu}, in the case that $(\delta\;\ell\;p) = (2\;0\;0)$, that the reduced OPE coefficient that vanishes at leading order, also vanishes at the $f_2$ order.  
It would be very interesting to carry out similar calculations for the other (KK-towers of) states studied in this paper.
However, in the case of the 2 states with $(\delta\;\ell\;p) = (2\;0\;0)$, it is known~\cite{Caron-Huot:2022sdy}, that \emph{both} the leading order OPE coefficients at \emph{weak} coupling are of the same order, \textit{i.e.} one is not subleading to the other.
It would also be interesting to see at which order, the vanishing reduced OPE coefficients begin to ``reappear''.

Finally, the sub-leading Casimir $j_1$ has played an important role in our analysis.
In general, the quadratic Casimir has proved to be a useful tool in CFT-data analysis. In~\cite{Gromov:2023hzc}, it was used to argue that all states in planar 4D $\mathcal{N} = 4$ SYM have the same constant integer shift in their dimension. 
Furthermore, it was conjectured that the sub-leading Casimir is constant on a KK-tower, and this was used to obtain a prediction of $d_1$ for \emph{all} states in a given KK-tower. 
Our main results rely on this conjecture. 
In the present paper, we showed that $j_1$ can be used as an elegant way to package constraints on the CFT-data of an entire KK-tower of states, in a $p$-independent way.
An important future direction is to prove the conjecture~\eqref{eqn:CasConj} of~\cite{Gromov:2023hzc}.
It would also be enlightening to see if further sub-leading orders of the Casimir are also able to repackage the CFT-data of planar 4D $\mathcal{N} = 4$ SYM at higher strong coupling orders, and whether this could be used to extract new predictions in the same spirit as this paper.

\begin{acknowledgments}
We are grateful to Luis Fernando Alday, Benjamin Basso, Nikolay Gromov, Dileep Jatkar, Alok Laddha, Jeremy Mann and Joao Silva for valuable discussions. The work of NS is supported by the European Research Council (ERC) under the European Union’s Horizon 2020 research and innovation programme (grant agreement No. 865075) EXACTC. JJ expresses gratitude to the people of India for their continuing support towards the study of basic sciences. 
\end{acknowledgments}

\appendix

\section{Relevant expressions}
The quadratic Casimir of $\mathrm{PSU}(2,2|4)$ is
\begin{multline}\label{eqn:Casimir}
J^2 = \frac{1}{2}(\Delta+2)^2 - 2 + \frac{1}{4} \ell_1 (\ell_1 + 2) + \frac{1}{4} \ell_2 (\ell_2 + 2) - \frac{1}{4} q_1 (q_1 + 2)- \frac{1}{4} q_2 (q_2 + 2) \\ - \frac{1}{8}(2p + q_1 + q_2)^2 - (2p + q_1 + q_2)\;.
\end{multline}
The average of the leading order OPE coefficient $\langle f_0\rangle$, depends only on $\delta$ and $\ell$ as shown in~\cite{Fardelli:2023fyq}. For the first three even-spin Regge trajectories, it is given by~\cite{Alday:2022uxp,Fardelli:2023fyq}
\begin{align}
    \langle f_0 \rangle_{\ell = 2(\delta -1)} &= \frac{r_0(\delta)}{\delta}\;, \label{eqn:f0t1} \\
    \langle f_0 \rangle_{\ell = 2(\delta - 2)} &= \frac{r_1(\delta)}{3}\left(2 \delta^2+3 \delta-8\right) \;, \label{eqn:f0t2} \\
    \langle f_0 \rangle_{\ell = 2(\delta - 3)} &= \frac{r_2(\delta)}{45}\left(10 \delta^4+43 \delta^3+8 \delta^2-352 \delta-192\right)\;. \label{eqn:f0t3}
\end{align}
Here $r_n$ is defined as~\cite{Alday:2022uxp}
\begin{align}\label{eqn:rndef}
    r_n(\delta) \equiv \frac{4^{2-2 \delta} \delta^{2 \delta-2 n-1}(2 \delta-2 n-1)}{\Gamma(\delta) \Gamma\left(\delta-\left\lfloor\frac{n}{2}\right\rfloor\right)}
    \;. 
\end{align}

\section{Explicit average formulas involving the sub-leading Casimir}
For $\langle f_0\,j_1\rangle$, on the first three even-spin Regge trajectories, we have
\begin{align}\label{eqn:f0j1t1}
	\langle f_0\,j_1\rangle_{\ell = 2 (\delta - 1)} &= {r_0(\delta)}\,
	\big(5\,\delta - 3\big)
	 \;, \\
\begin{split}\label{eqn:f0j1t2}
	\langle f_0\,j_1\rangle_{\ell = 2 (\delta - 2)} &= \frac{r_1(\delta)}{9}\, 
	\big(30\, \delta ^4+7\, \delta ^3 
    -147\, \delta ^2+212\, \delta -120\big)
	\;, 
\end{split}\\
\begin{split}\label{eqn:f0j1t3}
	\langle f_0\,j_1\rangle_{\ell = 2 (\delta - 3)} &= \frac{r_2(\delta)}{675}\,
	\big( 750\, \delta ^6+1775\, \delta ^5 -3667\, \delta^4 -18092\, \delta ^3 
	+45688\, \delta ^2 \\ &-59712\, \delta 
    -40320\, \big)
	\;,
\end{split}
\end{align}
where $r_n(\delta)$ is defined in equation~\eqref{eqn:rndef}.
Similarly, on the first three even-spin Regge trajectories, the explicit expressions for $\langle f_0\,j_1^2\rangle$ are
\begin{align}\label{eqn:f0j1sqt1}
	\langle f_0\,j_1^2\rangle_{\ell = 2 (\delta - 1)} &= 
     r_{0}(\delta)\,\delta\,\big(5\,\delta - 3\big)^2
     \;, \\
     \label{eqn:f0j1sqt2}
         \langle f_0\,j_1^2\rangle_{\ell = 2 (\delta - 2)} &= \frac{r_1(\delta)}{27} \big(
         450 \delta ^6-465 \delta ^5-1888 \delta ^4 
         +6663 \delta ^3-9248 \delta ^2+6180 \delta -1800 \big)
         \;, 
     \\
    \begin{split}
    \label{eqn:f0j1sqt3}
        \langle f_0\,j_1^2\rangle_{\ell = 2 (\delta - 3)} &= 
        \frac{r_2(\delta)}{10125}
        \big(56250 \delta ^8+24375 \delta ^7 
        -384800 \delta ^6-310397 \delta ^5+5867708 \delta ^4
       \\
        &
        -18469612 \delta ^3
        +21402888 \delta ^2-11053440 \delta 
        -8467200 \big)\;.
    \end{split}
\end{align}

\section{Predicted average formulas for $\langle f_0 \,d_1^2\rangle$ with non-zero $R$-charge}
Combining the equations~\eqref{eqn:f0j1sqt1} --~\eqref{eqn:f0j1sqt3} with the conjecture~\eqref{eqn:CasConj}, we can extract an average formula for $\langle f_0\,d_1^2\rangle$ when the $R$-charge $p>0$. For the first three Regge trajectories, we have
\begin{align}\label{eqn:f0d1sqt1}
	\langle f_0\,d_1^2\rangle_{\ell = 2 (\delta - 1)} &= 
     \frac{r_{0}(\delta)}{16\,\delta}\,
     \big[
     p^2+4 p+6 \delta ^2-2 \delta +4\big]^2
     \;, \\
     \begin{split}
     \label{eqn:f0d1sqt2}
         \langle f_0\,d_1^2\rangle_{\ell = 2 (\delta - 2)} &= \frac{r_1(\delta)}{432} \big[
         648 \delta ^6+828 \delta ^5-1504 \delta ^4+4908 \delta ^3-8624 \delta ^2+4608 \delta +p^4 \left(18 \delta ^2
         \right. \\ 
        &+ \left.
         27 \delta -72\right)
         + p^3 \left(144 \delta ^2+216 \delta -576\right)+p^2 \left(216 \delta ^4+300 \delta ^3-396 \delta ^2
         \right. \\ 
        &+ \left.
         1032 \delta -2016\right)
         +p \left(864 \delta ^4+1200 \delta ^3-2736 \delta ^2+2400 \delta -3456\right)\\&-2592
         \big]
         \;, 
     \end{split} 
\end{align}
\begin{align}
    \begin{split}
    \label{eqn:f0d1sqt3}
        \langle f_0\,d_1^2\rangle_{\ell = 2 (\delta - 3)} &= 
        \frac{r_2(\delta)}{162000}
        \big[
        81000 \delta ^8+366300 \delta ^7+311680 \delta ^6-1367108 \delta ^5+1344512 \delta ^4
        \\&-18021568 \delta ^3+8696352 \delta ^2
        -18063360 \delta +p^4 \left(2250 \delta ^4+9675 \delta ^3+1800 \delta ^2
        \right. \\ 
        &- \left.
        79200 \delta -43200\right)+p^3 \left(18000 \delta ^4+77400 \delta ^3
         14400 \delta ^2-633600 \delta -345600\right) \\&+ p^2 \left(27000 \delta ^6+119100 \delta ^5+98580 \delta ^4-612120 \delta ^3
        124320 \delta ^2-3409920 \delta 
        \right. \\ 
        &- \left.
        1382400\right)+p \left(108000 \delta ^6+476400 \delta ^5+250320 \delta ^4
        -3067680 \delta ^3
        \right. \\ 
        &- \left.
        612480 \delta ^2-8570880 \delta -2764800\right)-2764800
        \big]\;.
    \end{split}
\end{align}
These formulas are a prediction, which should be checked by a first-principles' derivation generalising the methods of~\cite{Fardelli:2023fyq} to the next order.

\section{Summary of perturbative OPE coefficient-data at strong coupling }

In the tables below, we collect all the perturbative OPE coefficient-data at strong coupling available in the literature, to the best of our knowledge, and include our results where applicable. One should keep in mind that spectral results are obtained by high-precision numerical fits of QSC-data, while drawing any conclusions from the data presented.

\setlength\LTleft{-25mm}
\setlength\LTright{-25mm}
\begin{xltabular}{\textwidth}{C|C|C|C|C|C|C|C|C|C|C|C|C}
 \text{St. No.} & \texttt{State ID} & \delta &
 \ell & \text{Traj.} & p  & d_1 & j_1 & \mathrm{KK}
 & f_0 & f_{t;2} & f_{t;4} & \text{Degs.}
 \\
 \midrule\midrule
 {1} & \text{}_ 2 \text{[0 0 1 1 1 1 0 0]}_ 1 & {1} & {0} & {1} & 0 & {2} & {2} & [0\;0] & {1\text{~\cite{Costa:2012cb}}} & {2 \zeta_3 + \frac{405}{32}\text{~\cite{Alday:2022xwz}}}
  & \text{see}\text{~\cite{Alday:2023mvu}} &  \\
 2 & \text{}_ 3 \text{[0 0 2 2 1 1 0 0]}_ 1 & 1 & 0 & 1 & 1 & \frac{13}{4} & 2 & [0\;0] & 1\text{~\cite{Fardelli:2023fyq}} & 2\zeta_3 + \frac{829}{32}\text{~\cite{Fardelli:2023fyq}} & \text{} & \text{} \\
 7 & \text{}_ 4 \text{[0 0 3 3 1 1 0 0]}_ 2 & 1 & 0 & 1 & 2 & 5 & 2 & [0\;0] & 1\text{~\cite{Fardelli:2023fyq}} & 2\zeta_3 + \frac{1733}{32}\text{~\cite{Fardelli:2023fyq}} & \text{} & \text{} \\
 23 & \text{}_ 5 \text{[0 0 4 4 1 1 0 0]}_ 1 & 1 & 0 & 1 & 3 & \frac{29}{4} & 2 & [0\;0] & 1\text{~\cite{Fardelli:2023fyq}} & 2\zeta_3 + \frac{3437}{32}\text{~\cite{Fardelli:2023fyq}} & \text{} & \text{} \\
 118 & \text{}_ 6 \text{[0 0 5 5 1 1 0 0]}_ 1 & 1 & 0 & 1 & 4 & 10 & 2 & [0\;0] & 1\text{~\cite{Fardelli:2023fyq}} & 2\zeta_3 + \frac{6357}{32}\text{~\cite{Fardelli:2023fyq}} & \text{} & \text{} \\
\midrule
 {12} & \text{}_ 4 \text{[0 2 1 1 1 1 2 0]}_ 1 & {2} & {2} & {1} & {0} & {6} & {14} & \mathrm{[0\;0]} & \frac{3}{4}\text{~\cite{Alday:2022uxp}} & 6 \zeta_3 +\frac{5703}{256}\text{~\cite{Alday:2022xwz}} &  
 \text{see}\text{~\cite{Alday:2023mvu}}
 &  \\
 39 & \text{}_ 5 \text{[0 2 2 2 1 1 2 0]}_ 1 & 2 & 2 & 1 & 1 & \frac{29}{4} & 14 & \mathrm{[0\;0]} & \frac{3}{4}\text{~\cite{Fardelli:2023fyq}} & 6 \zeta_3 +\frac{11199}{256}\text{~\cite{Fardelli:2023fyq}} & \text{} & \text{} \\
 206 & \text{}_ 6 \text{[0 2 3 3 1 1 2 0]}_ 3 & 2 & 2 & 1 & 2 & 9 & 14 & \mathrm{[0\;0]} & \frac{3}{4}\text{~\cite{Fardelli:2023fyq}} & 6 \zeta_3 +\frac{20343}{256}\text{~\cite{Fardelli:2023fyq}} & \text{} & \text{} \\
 \midrule
 {219} & \text{}_ 6 \text{[0 4 1 1 1 1 4 0]}_ 1 & {3} & {4} & {1} & {0} & {13} & {36} & [0\;0] & \frac{405}{1024}\text{~\cite{Alday:2022uxp}} & 
 \frac{3645}{512}\zeta_3 + \frac{566595}{32768}\text{~\cite{Alday:2022xwz}} & 
 \text{see}\text{~\cite{Alday:2023mvu}}
 & \text{} \\
 \midrule
 3 & \text{}_ 4 \text{[0 0 2 2 2 2 0 0]}_ 1 & 2 & 0 & 2 & 0 & 2 & 2 & [0\;0]_1 & 0\text{~\cite{Gromov:2023hzc,Alday:2023mvu}} & 0\text{~\cite{Alday:2023mvu}} & \text{} & \text{} \\
 4 & \text{}_ 4 \text{[0 0 2 2 2 2 0 0]}_ 2 & 2 & 0 & 2 & 0 & 8 & 14 & [0\;0]_2 & \frac{1}{4}\text{~\cite{Gromov:2023hzc,Alday:2023mvu}} & 2\zeta_3 - \frac{387}{256}\text{~\cite{Alday:2023mvu}} & \text{} & \text{} \\
 17 & \text{}_ 5 \text{[0 0 3 3 2 2 0 0]}_ 1 & 2 & 0 & 2 & 1 & \frac{13}{4} & 2 & [0\;0]_1 & {\color{blue} 0} & \text{} & \text{} & \text{} \\
 18 & \text{}_ 5 \text{[0 0 3 3 2 2 0 0]}_ 2 & 2 & 0 & 2 & 1 & \frac{37}{4} & 14 & [0\;0]_2 & {\color{blue} \frac{1}{4}} & \text{} & \text{} & \text{} \\
 6 & \text{}_ 4 \text{[0 0 3 3 1 1 0 0]}_ 1 & 2 & 0 & 2 & 2 & 5 & 2 & [2\;0] & {\color{blue} 0} & \text{} & \text{} & \text{} \\
 107 & \text{}_ 6 \text{[0 0 4 4 2 2 0 0]}_ 3 & 2 & 0 & 2 & 2 & 11 & 14  & [0\;0]_2 & {\color{blue} \frac{1}{4}} & \text{} & \text{} & \text{} \\
 109 & \text{}_ 6 \text{[0 0 4 4 2 2 0 0]}_ 5 & 2 & 0 & 2 & 2 & 5 & 2 & [0\;0] & {\color{blue} 0} & \text{} & \text{} & \text{} \\
 24 & \text{}_ 5 \text{[0 0 4 4 1 1 0 0]}_ 2 & 2 & 0 & 2 & 3 & \frac{29}{4} & 2 & [2\;0] &  {\color{blue} 0} & \text{} & \text{} & \text{} \\
 119 & \text{}_ 6 \text{[0 0 5 5 1 1 0 0]}_ 2 & 2 & 0 & 2 & 4 & 10 & 2 & [2\;0] & {\color{blue} 0} & \text{} & \text{} & \text{} \\
 \midrule
 196 & \text{}_ 6 \text{[0 2 2 2 2 2 2 0]}_ 2 & 3 & 2 & 2 & 0 & 8 & 18 & [0\;0]_1 & {\color{blue} 0} & \text{} & \text{} & \text{} \\
 197 & \text{}_ 6 \text{[0 2 2 2 2 2 2 0]}_ 3 & 3 & 2 & 2 & 0 & 17 & 36 & [0\;0]_2 & {\color{blue} \frac{243}{1024}} & \text{} & \text{} & \text{} \\
 198 & \text{}_ 6 \text{[0 2 2 2 2 2 2 0]}_ 4 & 3 & 2 & 2 & 0 & 13 & 28 & [0\;0]_3 & {\color{blue} \frac{135}{1024}} & \text{} & \text{} & 199 \\
 199 & \text{}_ 6 \text{[0 2 2 2 2 2 2 0]}_ 5 & 3 & 2 & 2 & 0 & 13 & 28 & [0\;0]_4 & {\color{blue} \frac{135}{1024}} & \text{} & \text{} & 198 \\
 205 & \text{}_ 6 \text{[0 2 3 3 1 1 2 0]}_ 2 & 3 & 2 & 2 & 2 & 11 & 18 & [0\;0]_1 & {\color{blue} 0} & \text{} & \text{} & \text{} \\
 207 & \text{}_ 6 \text{[0 2 3 3 1 1 2 0]}_ 4 & 3 & 2 & 2 & 2 & 11 & 18 & [2\;0]_1 & {\color{blue} 0} & \text{} & \text{} & 208 \\
 208 & \text{}_ 6 \text{[0 2 3 3 1 1 2 0]}_ 5 & 3 & 2 & 2 & 2 & 11 & 18 & [2\;0]_2 & {\color{blue} 0} & \text{} & \text{} & 207 \\
 \midrule
 76 & \text{}_ 6 \text{[0 0 2 2 2 2 1 1]}_ 1 & 3 & 0 & 3 & 0 & \frac{31}{4} & \frac{27}{2} & [0\;0]_1 & {\color{blue} 0} & \text{} & \text{} & 85 \\
 78 & \text{}_ 6 \text{[0 0 3 3 3 3 0 0]}_ 1 & 3 & 0 & 3 & 0 & 1 & 0 & [0\;0]_2 & {\color{blue} 0} & \text{} & \text{} & \text{} \\
 81 & \text{}_ 6 \text{[0 0 3 3 3 3 0 0]}_ 4 & 3 & 0 & 3 & 0 & 10 & 18 & [0\;0]_3 & {\color{blue} 0} & \text{} & \text{} & \text{} \\
 83 & \text{}_ 6 \text{[1 1 2 2 2 2 1 1]}_ 1 & 3 & 0 & 3 & 0 & 11 & 20 & [0\;0]_4 & {\color{blue} \frac{25}{1024}} & \text{} & \text{} & \text{} \\
 84 & \text{}_ 6 \text{[1 1 2 2 2 2 1 1]}_ 2 & 3 & 0 & 3 & 0 & 19 & 36 & [0\;0]_5 & {\color{blue} \frac{81}{1024}} & \text{} & \text{} & \text{} \\
 85 & \text{}_ 6 \text{[1 1 3 3 3 3 0 0]}_ 1 & 3 & 0 & 3 & 0 & \frac{31}{4} & \frac{27}{2} & [0\;0]_6 & {\color{blue} 0} & \text{} & \text{} & 76 \\
 19 & \text{}_ 5 \text{[0 0 3 3 2 2 0 0]}_ 3 & 3 & 0 & 3 & 1 & \text{} &  & [0\;0]_1 &  {\color{blue} 0} & \text{} & \text{} & 20 \\
 20 & \text{}_ 5 \text{[0 0 3 3 2 2 0 0]}_ 4 & 3 & 0 & 3 & 1 & \text{} & & [0\;0]_6 & {\color{blue} 0} & \text{} & \text{} & 19 \\
 105 & \text{}_ 6 \text{[0 0 4 4 2 2 0 0]}_ 1 & 3 & 0 & 3 & 2 & \text{} & & [0\;0]_1 & {\color{blue} 0} & \text{} & \text{} & 106 \\
 106 & \text{}_ 6 \text{[0 0 4 4 2 2 0 0]}_ 2 & 3 & 0 & 3 & 2 & \text{} & & [0\;0]_6 & {\color{blue} 0} & \text{} & \text{} & 105 \\
 108 & \text{}_ 6 \text{[0 0 4 4 2 2 0 0]}_ 4 & 3 & 0 & 3 & 2 & 13 & 18 & [2\;0]_1 & {\color{blue} 0} & \text{} & \text{} & \text{} \\
 112 & \text{}_ 6 \text{[0 0 4 4 2 2 0 0]}_ 8 & 3 & 0 & 3 & 2 & 4 & 0 & [2\;0]_2 & {\color{blue} 0} & \text{} & \text{} & \text{} \\
 113 & \text{}_ 6 \text{[0 0 4 4 2 2 0 0]}_ 9 & 3 & 0 & 3 & 2 & 9 & 10 & [2\;0]_3 & {\color{blue} 0} & \text{} & \text{} & 114 \\
 114 & \text{}_ 6 \text{[0 0 4 4 2 2 0 0]}_{10} & 3 & 0 & 3 & 2 & 9 & 10 & [2\;0]_4 & {\color{blue} 0} & \text{} & \text{} & 113 \\
 120 & \text{}_ 6 \text{[0 0 5 5 1 1 0 0]}_ 3 & 3 & 0 & 3 & 4 & 9 & 0 & [4\;0] & {\color{blue} 0} & \text{} & \text{} & \text{} \\
    \caption{
    \label{tab:EvenReggeStates} 
Perturbative CFT-data for the 40 lowest lying states in planar 4D $\mathcal{N} = 4$ SYM on even-spin Regge trajectories: whose Lorentz spin labels are of the form $[\ell\;\ell]$ and $R$-symmetry labels are like $[0\;p\;0]$.
For every state, we display its State Number (St. No.) and \texttt{State ID}, both of which are unique identifiers of a given state, introduced in~\cite{Gromov:2023hzc}. We also display the string mass level $\delta$, spin $\ell$ , Regge trajectory number,  $R$-charge $p$, sub-sub-leading dimension $d_1$ from~\cite{Gromov:2023hzc}, sub-leading Casimir $j_1$ from~\cite{Gromov:2023hzc}, the KK-tower assigned in~\cite{Gromov:2023hzc}, and exact degeneracies (see~\cite{Gromov:2023hzc}) of the state. Finally we present the strong coupling expansion coefficients of the OPE coefficient of a state. We have added references to the available results in the literature. New results obtained by us are coloured blue.
    }
\end{xltabular}

\setlength\LTleft{+0mm}
\setlength\LTright{+0mm}
\begin{xltabular}{\textwidth}{C|C|C|C|C|C|C|C|C|C|C}
 \text{St. No.} & \texttt{State ID} & \delta &
 \ell & \text{Traj.} & p  & d_1 & j_1 & \mathrm{KK} & f_{t;2} & \text{Degs.}
 \\
 \midrule\midrule
 9 & \text{}_ 4 \text{[0 1 2 2 1 1 1 0]}_ 1 & 2 & 1 & 1 & 1 & \frac{11}{2} & 8 & [1\;0]_1 & \frac{5}{4}\text{~\cite{Fardelli:2023fyq,Gromov:2023hzc}} & 10 \\
 10 & \text{}_ 4 \text{[0 1 2 2 1 1 1 0]}_ 2 & 2 & 1 & 1 & 1 & \frac{11}{2} & 8 & [1\;0]_2 & \frac{5}{4}\text{~\cite{Fardelli:2023fyq,Gromov:2023hzc}} & 9 \\
 34 & \text{}_ 5 \text{[0 1 3 3 1 1 1 0]}_ 1 & 2 & 1 & 1 & 2 & \frac{29}{4} & 8 & [1\;0]_1 & 3\text{~\cite{Fardelli:2023fyq,Gromov:2023hzc}} & 35 \\
 35 & \text{}_ 5 \text{[0 1 3 3 1 1 1 0]}_ 2 & 2 & 1 & 1 & 2 & \frac{29}{4} & 8 & [1\;0]_2 & 3\text{~\cite{Fardelli:2023fyq,Gromov:2023hzc}} & 34 \\
 173 & \text{}_ 6 \text{[0 1 4 4 1 1 1 0]}_ 3 & 2 & 1 & 1 & 3 & \frac{19}{2} & 8 & [1\;0]_1 & \frac{21}{4}\text{~\cite{Fardelli:2023fyq,Gromov:2023hzc}} & 174 \\
 174 & \text{}_ 6 \text{[0 1 4 4 1 1 1 0]}_ 4 & 2 & 1 & 1 & 3 & \frac{19}{2} & 8 & [1\;0]_2 & \frac{21}{4}\text{~\cite{Fardelli:2023fyq,Gromov:2023hzc}} & 173 \\
\midrule
 214 & \text{}_ 6 \text{[0 3 2 2 1 1 3 0]}_ 1 & 3 & 3 & 1 & 1 & 12 & 27 & [1\;0]_1 & \frac{135}{128}\text{~\cite{Fardelli:2023fyq,Gromov:2023hzc}}  & 215 \\
 215 & \text{}_ 6 \text{[0 3 2 2 1 1 3 0]}_ 2 & 3 & 3 & 1 & 1 & 12 & 27 & [1\;0]_2 & \frac{135}{128}\text{~\cite{Fardelli:2023fyq,Gromov:2023hzc}}  & 214 \\
\caption{\label{tab:OddReggeStates} 
Perturbative CFT-data for the 8 lowest lying states in planar 4D $\mathcal{N} = 4$ SYM on odd-spin Regge trajectories: whose Lorentz spin labels are of the form $[\ell\;\ell]$ and $R$-symmetry labels are like $[0\;p\;0]$.
For every state, we display its St. No. and \texttt{State ID}, string mass level $\delta$, spin $\ell$ , Regge trajectory number,  $R$-charge $p$, sub-sub-leading dimension $d_1$ from~\cite{Gromov:2023hzc}, sub-leading Casimir $j_1$ from~\cite{Gromov:2023hzc}, the KK-tower assigned in~\cite{Gromov:2023hzc}, and exact degeneracies (see~\cite{Gromov:2023hzc}) of the state. 
All the states are on the leading odd-spin Regge trajectory, \textit{i.e.}, for these states $\ell = 2(\delta - 1) - 1 = 2\,\delta- 3$. The degeneracy of all states on this Regge trajectory is 2~\cite{Alday:2023flc}. Furthermore, in all the states considered in~\cite{Gromov:2023hzc}, it was observed that the degeneracy is exact, \textit{i.e.} that the scaling dimensions of the 2 degenerate states are indistinguishable. In particular, this is true to all orders in perturbation theory. Therefore, the average formula for $\langle f_{t;2}\rangle$ obtained by~\cite{Fardelli:2023fyq} is actually a prediction for the precise OPE coefficient. We also display this prediction.}
\end{xltabular}

\bibliographystyle{JHEP.bst}
\bibliography{references}

\providecommand{\href}[2]{#2}\begingroup\raggedright\begin{thebibliography}{10}

\bibitem{Rattazzi:2008pe}
R.~Rattazzi, V.~S. Rychkov, E.~Tonni, and A.~Vichi, {\it {Bounding scalar
  operator dimensions in 4D CFT}},  {\em JHEP} {\bf 12} (2008) 031,
  [\href{http://arxiv.org/abs/0807.0004}{{\tt arXiv:0807.0004}}].

\bibitem{El-Showk:2012cjh}
S.~El-Showk, M.~F. Paulos, D.~Poland, S.~Rychkov, D.~Simmons-Duffin, and
  A.~Vichi, {\it {Solving the 3D Ising Model with the Conformal Bootstrap}},
  {\em Phys. Rev. D} {\bf 86} (2012) 025022,
  [\href{http://arxiv.org/abs/1203.6064}{{\tt arXiv:1203.6064}}].

\bibitem{Kos:2016ysd}
F.~Kos, D.~Poland, D.~Simmons-Duffin, and A.~Vichi, {\it {Precision Islands in
  the Ising and $O(N)$ Models}},  {\em JHEP} {\bf 08} (2016) 036,
  [\href{http://arxiv.org/abs/1603.04436}{{\tt arXiv:1603.04436}}].

\bibitem{Gromov:2009tv}
N.~Gromov, V.~Kazakov, and P.~Vieira, {\it {Exact Spectrum of Anomalous
  Dimensions of Planar N=4 Supersymmetric Yang-Mills Theory}},  {\em Phys. Rev.
  Lett.} {\bf 103} (2009) 131601, [\href{http://arxiv.org/abs/0901.3753}{{\tt
  arXiv:0901.3753}}].

\bibitem{Gromov:2013pga}
N.~Gromov, V.~Kazakov, S.~Leurent, and D.~Volin, {\it {Quantum Spectral Curve
  for Planar $\mathcal{N} = 4$ Super-Yang-Mills Theory}},  {\em Phys. Rev.
  Lett.} {\bf 112} (2014), no.~1 011602,
  [\href{http://arxiv.org/abs/1305.1939}{{\tt arXiv:1305.1939}}].

\bibitem{Cavaglia:2014exa}
A.~Cavagli\`a, D.~Fioravanti, N.~Gromov, and R.~Tateo, {\it {Quantum Spectral
  Curve of the $\mathcal N=$ 6 Supersymmetric Chern-Simons Theory}},  {\em
  Phys. Rev. Lett.} {\bf 113} (2014), no.~2 021601,
  [\href{http://arxiv.org/abs/1403.1859}{{\tt arXiv:1403.1859}}].

\bibitem{Basso:2015zoa}
B.~Basso, S.~Komatsu, and P.~Vieira, ``{Structure Constants and Integrable
  Bootstrap in Planar N=4 SYM Theory}.'' 5, 2015.

\bibitem{Cavaglia:2018lxi}
A.~Cavagli{\`a}, N.~Gromov, and F.~Levkovich-Maslyuk, {\it {Quantum spectral
  curve and structure constants in $ \mathcal{N}=4 $ SYM: cusps in the ladder
  limit}},  {\em JHEP} {\bf 10} (2018) 060,
  [\href{http://arxiv.org/abs/1802.04237}{{\tt arXiv:1802.04237}}].

\bibitem{Jiang:2019zig}
Y.~Jiang, S.~Komatsu, and E.~Vescovi, {\it {Exact Three-Point Functions of
  Determinant Operators in Planar $N=4$ Supersymmetric Yang-Mills Theory}},
  {\em Phys. Rev. Lett.} {\bf 123} (2019), no.~19 191601,
  [\href{http://arxiv.org/abs/1907.11242}{{\tt arXiv:1907.11242}}].

\bibitem{Cavaglia:2021mft}
A.~Cavagli{\`a}, N.~Gromov, and F.~Levkovich-Maslyuk, {\it {Separation of
  variables in AdS/CFT: functional approach for the fishnet CFT}},  {\em JHEP}
  {\bf 06} (2021) 131, [\href{http://arxiv.org/abs/2103.15800}{{\tt
  arXiv:2103.15800}}].

\bibitem{Grabner:2020nis}
D.~Grabner, N.~Gromov, and J.~Julius, {\it {Excited States of One-Dimensional
  Defect CFTs from the Quantum Spectral Curve}},  {\em JHEP} {\bf 07} (2020)
  042, [\href{http://arxiv.org/abs/2001.11039}{{\tt arXiv:2001.11039}}].

\bibitem{Julius:2021uka}
J.~Julius, {\em {Modern techniques for solvable models}}.
\newblock PhD thesis, King's Coll. London, 2021.

\bibitem{Bercini:2022jxo}
C.~Bercini, A.~Homrich, and P.~Vieira, {\it {Structure Constants in
  $\mathcal{N} = 4$ SYM and Separation of Variables}},
  \href{http://arxiv.org/abs/2210.04923}{{\tt arXiv:2210.04923}}.

\bibitem{Cavaglia:2021eqr}
A.~Cavagli\`a, N.~Gromov, B.~Stefa\'nski, Jr., Jr., and A.~Torrielli, {\it
  {Quantum Spectral Curve for AdS$_{3}$/CFT$_{2}$: a proposal}},  {\em JHEP}
  {\bf 12} (2021) 048, [\href{http://arxiv.org/abs/2109.05500}{{\tt
  arXiv:2109.05500}}].

\bibitem{Ekhammar:2021pys}
S.~Ekhammar and D.~Volin, {\it {Monodromy bootstrap for SU(2|2) quantum
  spectral curves: from Hubbard model to AdS$_{3}$/CFT$_{2}$}},  {\em JHEP}
  {\bf 03} (2022) 192, [\href{http://arxiv.org/abs/2109.06164}{{\tt
  arXiv:2109.06164}}].

\bibitem{Cavaglia:2021bnz}
A.~Cavagli\`a, N.~Gromov, J.~Julius, and M.~Preti, {\it {Integrability and
  conformal bootstrap: One dimensional defect conformal field theory}},  {\em
  Phys. Rev. D} {\bf 105} (2022), no.~2 L021902,
  [\href{http://arxiv.org/abs/2107.08510}{{\tt arXiv:2107.08510}}].

\bibitem{Cavaglia:2022qpg}
A.~Cavagli\`a, N.~Gromov, J.~Julius, and M.~Preti, {\it {Bootstrability in
  Defect CFT: Integrated Correlators and Sharper Bounds}},
  \href{http://arxiv.org/abs/2203.09556}{{\tt arXiv:2203.09556}}.

\bibitem{Cavaglia:2022yvv}
A.~Cavagli\`a, N.~Gromov, J.~Julius, and M.~Preti, {\it {Integrated correlators
  from integrability: Maldacena-Wilson line in $ \mathcal{N} $ = 4 SYM}},  {\em
  JHEP} {\bf 04} (2023) 026, [\href{http://arxiv.org/abs/2211.03203}{{\tt
  arXiv:2211.03203}}].

\bibitem{Caron-Huot:2022sdy}
S.~Caron-Huot, F.~Coronado, A.-K. Trinh, and Z.~Zahraee, {\it {Bootstrapping
  $\mathcal{N}=4$ sYM correlators using integrability}},
  \href{http://arxiv.org/abs/2207.01615}{{\tt arXiv:2207.01615}}.

\bibitem{Niarchos:2023lot}
V.~Niarchos, C.~Papageorgakis, P.~Richmond, A.~G. Stapleton, and M.~Woolley,
  {\it {Bootstrability in Line-Defect CFT with Improved Truncation Methods}},
  \href{http://arxiv.org/abs/2306.15730}{{\tt arXiv:2306.15730}}.

\bibitem{Gromov:2015wca}
N.~Gromov, F.~Levkovich-Maslyuk, and G.~Sizov, {\it {Quantum Spectral Curve and
  the Numerical Solution of the Spectral Problem in AdS5/CFT4}},  {\em JHEP}
  {\bf 06} (2016) 036, [\href{http://arxiv.org/abs/1504.06640}{{\tt
  arXiv:1504.06640}}].

\bibitem{Gromov:2023hzc}
N.~Gromov, A.~Hegedus, J.~Julius, and N.~Sokolova, {\it {Fast QSC Solver: tool
  for systematic study of N=4 Super-Yang-Mills spectrum}},
  \href{http://arxiv.org/abs/2306.12379}{{\tt arXiv:2306.12379}}.

\bibitem{Alday:2022uxp}
L.~F. Alday, T.~Hansen, and J.~A. Silva, {\it {AdS Virasoro-Shapiro from
  dispersive sum rules}},  {\em JHEP} {\bf 10} (2022) 036,
  [\href{http://arxiv.org/abs/2204.07542}{{\tt arXiv:2204.07542}}].

\bibitem{Alday:2022xwz}
L.~F. Alday, T.~Hansen, and J.~A. Silva, {\it {AdS Virasoro-Shapiro from
  single-valued periods}},  \href{http://arxiv.org/abs/2209.06223}{{\tt
  arXiv:2209.06223}}.

\bibitem{Alday:2023mvu}
L.~F. Alday and T.~Hansen, {\it {The AdS Virasoro-Shapiro Amplitude}},
  \href{http://arxiv.org/abs/2306.12786}{{\tt arXiv:2306.12786}}.

\bibitem{Fardelli:2023fyq}
G.~Fardelli, T.~Hansen, and J.~A. Silva, {\it {AdS Virasoro-Shapiro amplitude
  with KK modes}},  \href{http://arxiv.org/abs/2308.03683}{{\tt
  arXiv:2308.03683}}.

\bibitem{Dolan:2001tt}
F.~A. Dolan and H.~Osborn, {\it {Superconformal symmetry, correlation functions
  and the operator product expansion}},  {\em Nucl. Phys. B} {\bf 629} (2002)
  3--73, [\href{http://arxiv.org/abs/hep-th/0112251}{{\tt hep-th/0112251}}].

\bibitem{Arutyunov:2002fh}
G.~Arutyunov, F.~A. Dolan, H.~Osborn, and E.~Sokatchev, {\it {Correlation
  functions and massive Kaluza-Klein modes in the AdS / CFT correspondence}},
  {\em Nucl. Phys. B} {\bf 665} (2003) 273--324,
  [\href{http://arxiv.org/abs/hep-th/0212116}{{\tt hep-th/0212116}}].

\bibitem{Dolan:2003hv}
F.~A. Dolan and H.~Osborn, {\it {Conformal partial waves and the operator
  product expansion}},  {\em Nucl. Phys.} {\bf B678} (2004) 491--507,
  [\href{http://arxiv.org/abs/hep-th/0309180}{{\tt hep-th/0309180}}].

\bibitem{Dolan:2004iy}
F.~A. Dolan and H.~Osborn, {\it {Conformal partial wave expansions for N=4
  chiral four point functions}},  {\em Annals Phys.} {\bf 321} (2006) 581--626,
  [\href{http://arxiv.org/abs/hep-th/0412335}{{\tt hep-th/0412335}}].

\bibitem{Caron-Huot:2018kta}
S.~Caron-Huot and A.-K. Trinh, {\it {All tree-level correlators in
  AdS$_{5}$\texttimes{}S$_{5}$ supergravity: hidden ten-dimensional conformal
  symmetry}},  {\em JHEP} {\bf 01} (2019) 196,
  [\href{http://arxiv.org/abs/1809.09173}{{\tt arXiv:1809.09173}}].

\bibitem{Gubser:1998bc}
S.~S. Gubser, I.~R. Klebanov, and A.~M. Polyakov, {\it {Gauge theory
  correlators from noncritical string theory}},  {\em Phys. Lett.} {\bf B428}
  (1998) 105--114, [\href{http://arxiv.org/abs/hep-th/9802109}{{\tt
  hep-th/9802109}}].

\bibitem{Gromov:2014caa}
N.~Gromov, V.~Kazakov, S.~Leurent, and D.~Volin, {\it {Quantum spectral curve
  for arbitrary state/operator in AdS$_{5}$/CFT$_{4}$}},  {\em JHEP} {\bf 09}
  (2015) 187, [\href{http://arxiv.org/abs/1405.4857}{{\tt arXiv:1405.4857}}].

\bibitem{Hegedus:2016eop}
A.~Heged\H{u}s and J.~Konczer, {\it {Strong coupling results in the AdS$_{5}$
  /CFT$_{4}$ correspondence from the numerical solution of the quantum spectral
  curve}},  {\em JHEP} {\bf 08} (2016) 061,
  [\href{http://arxiv.org/abs/1604.02346}{{\tt arXiv:1604.02346}}].

\bibitem{Alday:2023flc}
L.~F. Alday, T.~Hansen, and J.~A. Silva, {\it {On the spectrum and structure
  constants of short operators in N=4 SYM at strong coupling}},
  \href{http://arxiv.org/abs/2303.08834}{{\tt arXiv:2303.08834}}.

\bibitem{Bianchi:2003wx}
M.~Bianchi, J.~F. Morales, and H.~Samtleben, {\it {On stringy AdS(5) x S**5 and
  higher spin holography}},  {\em JHEP} {\bf 07} (2003) 062,
  [\href{http://arxiv.org/abs/hep-th/0305052}{{\tt hep-th/0305052}}].

\bibitem{Alday:2023jdk}
L.~F. Alday, T.~Hansen, and J.~A. Silva, {\it {Emergent world-sheet for the AdS
  Virasoro-Shapiro amplitude}},  \href{http://arxiv.org/abs/2305.03593}{{\tt
  arXiv:2305.03593}}.

\bibitem{Marboe:2017dmb}
C.~Marboe and D.~Volin, {\it {The full spectrum of AdS5/CFT4 I: Representation
  theory and one-loop Q-system}},  {\em J. Phys. A} {\bf 51} (2018), no.~16
  165401, [\href{http://arxiv.org/abs/1701.03704}{{\tt arXiv:1701.03704}}].

\bibitem{Marboe:2018ugv}
C.~Marboe and D.~Volin, {\it {The full spectrum of AdS$_5$/CFT$_4$ II: Weak
  coupling expansion via the quantum spectral curve}},  {\em J. Phys. A} {\bf
  54} (2021), no.~5 055201, [\href{http://arxiv.org/abs/1812.09238}{{\tt
  arXiv:1812.09238}}].

\bibitem{AHSPrivate}
L.~F. Alday, T.~Hansen, and J.~A. Silva, ``{\it Private communication}.''

\bibitem{Costa:2012cb}
M.~S. Costa, V.~Goncalves, and J.~Penedones, {\it {Conformal Regge theory}},
  {\em JHEP} {\bf 12} (2012) 091, [\href{http://arxiv.org/abs/1209.4355}{{\tt
  arXiv:1209.4355}}].

\bibitem{Goncalves:2014ffa}
V.~Gon\c{c}alves, {\it {Four point function of $\mathcal{N}=4$ stress-tensor
  multiplet at strong coupling}},  {\em JHEP} {\bf 04} (2015) 150,
  [\href{http://arxiv.org/abs/1411.1675}{{\tt arXiv:1411.1675}}].

\end{thebibliography}\endgroup

\end{document}